\documentclass[letterpaper]{JHEP}
\usepackage{epsfig}

\title{The Reduced Open Membrane Metric}
\author{Jan Pieter van der Schaar\\ 
	Michigan Center for Theoretical Physics\\
	Randall Laboratory, Department of Physics, University of Michigan\\
	Ann Arbor, MI 48109-1120, USA\\
	E-mail: \email{jpschaar@umich.edu}}
\abstract{We discuss the reduction of the open membrane metric and 
determine the (previously unknown) conformal factor. We also construct
$SL(2,R)$ invariant open string metrics and complex open string coupling 
constants by reducing the open membrane metric on a 2-torus. In doing so
we also clarify some issues on manifest $SL(2,R)$ symmetry of the D3-brane. 
We remark on the consequences of our results for the recently 
conjectured existence of decoupled $(p,q)$ non-commutative open string 
theories in type IIB string theory.}
 

\preprint{MCTP-01-26 \\ hep-th/0106046}

\begin{document}
\bibliographystyle{JHEP}

\section{Introduction}

A widespread interest in non-commutative theories from the perspective of 
string theory started with the paper of Seiberg and Witten 
\cite{Seiberg-Witten}.
They realized that, among other things, open strings probing the D-brane 
are most naturally described using open string moduli which depend on the
2-form ${\cal F}$ on the D-brane; the open string metric 
$(G^{-1}_{os})_{ab}$ 
and the open string coupling constant $\lambda_{os}$. 
As first discussed in 
\cite{Bergshoeff-NCM5}, this situation was generalized to include the 
M5-brane of M-theory, which involved a conjectured open membrane (OM) metric 
which depended on the nonlinearly self-dual 3-form ${\cal H}$ on the M5-brane.

In a somewhat different context it was shown in 
\cite{Gibbons-Herdeiro, Gibbons-West} that both the open string and 
the open membrane metric provide the propagation 
cone for all D3- and M5-brane degrees of freedom, which always lies 
inside the bulk Einstein light-cone. Consequently they also showed 
explicitly that all equations of motion on the M5-brane and the D3-brane 
can be conveniently written using a symmetric tensor that is in the same 
conformal class as the open string and open membrane (co-) metric, the 
so-called Boillat metric.

Attempts to decouple D-brane theories from the bulk supergravity 
were at first only successful for magnetic field strengths giving rise
to non-commutative gauge theories \cite{Seiberg-Witten}. 
To avoid problems with unitarity in spatio-temporal non-commutative 
field theory \cite{Seiberg-causal, Barbon-Rabinovici} one was lead to
the discovery of decoupled non-commutative open string theories (NCOS) on 
D-branes \cite{Gopakumar-NCOS, Seiberg-ncos}, forcing one to take a critical 
limit for the electric field strength. 
By considering a decoupling limit where the electric 3-form field strength 
on the M5-brane becomes (nearly) critical this lead to the 
introduction of OM-theory \cite{Gopakumar-OM, Bergshoeff-OM}, which 
can be understood as the mother of all (spatially and spatio-temporally) 
non-commutative theories.

In \cite{Bergshoeff-OM} the conformal factor in front of the open
membrane metric was fixed using a decoupling argument: the open 
membrane metric $(G^{-1}_{OM})_{ab}$ and the $D=11$ Planck length $\ell_p$ 
should define a finite
(non-commutative) length scale $\ell_g$ in the OM-theory decoupling limit. 
This allowed 
one to fix the conformal factor, but only up to terms that vanish in the 
decoupling limit.
In this paper we would like to determine the conformal factor
without ever considering a decoupling limit. Our guiding principle will
be that we want the open membrane metric 
$(G^{-1}_{OM})_{ab}$ to reduce to the open string metric 
$(G^{-1}_{os})_{ab}$ {\it and} the open string coupling constant 
$\lambda_{os}$. As we will see this will enable us, using reduction 
ansatze analogous to the bulk M-theory/IIA relations, to fix the 
conformal factor.

Reducing the open membrane metric to an expression depending only on the 2-form
${\cal F}$ is nontrivial because of the nonlinear self-duality equation
on the M5-brane and consequently, the nonlinear duality equation on 
the D4-brane (described by nonlinear Dirac-Born-Infeld (DBI) theory, for 
a review see \cite{Tseytlin-BIrev}). 
The reduction of the open membrane metric was performed in 
\cite{Howe1, Howe2}, but to obtain their 
result we will proceed in a slightly different (and less general) manner.
More importantly, they were unaware at that time of the interpretation of their
symmetric tensor as the open membrane metric and therefore did not
consider the internal component of that tensor upon reduction, 
which we will now interpret as the open string coupling constant. 
So keeping these newly acquired insights in mind and redoing their 
calculation \cite{Howe1, Howe2}, simplifying matters by imposing a 
constraint that will restrict our attention to rank 2 
solutions of the 2-form ${\cal F}$, we are able to fix the conformal 
factor of the open membrane metric as to give us both the open string 
metric and the open string coupling constant upon (double dimensional) 
reduction of the M5-brane. 
  
After fixing the conformal factor we 
reduce the open membrane metric on a 2-torus giving 
an $SL(2,R)$ invariant open string metric and introducing 
a complex open string coupling constant as the modular 
parameter of the torus as seen by the open membrane. 
We find that generically the situation on the D3-brane is very
similar to the M5-brane in the sense that the doublet of 
2-form field strengths has to satisfy a nonlinear self-duality 
equation. By $SL(2,R)$ rotating into a special frame
we can reproduce the Seiberg-Witten results, at the cost of 
giving up manifest $SL(2,R)$ invariance. These results are of course
intimately connected to previous investigations of $SL(2,R)$ invariance
on the D3-brane \cite{Gibbons-Rasheed, Tseytlin-BIsd, Berman-SLDBI, 
Berman-SLM5, Cederwall-SLD3}, but we will make full use of the open 
membrane metric idea in the sense that the $SL(2,R)$ symmetry is generated 
by the open membrane torus, instead
of the usual `bulk' torus. In that way we will  
also be connecting to work done with respect
to the $SL(2,R)$ symmetry of NCOS theories on the D3-brane 
\cite{Lu-Singh, Russo-Jabbari, Kawano-Terashima}. 
In the same section we also comment on the decoupling limit
that would give rise to $(p,q)$ non-commutative open strings that
were discussed recently in \cite{Gran-Nielsen,Lu}. We end with
some conclusions, remarks and possible future extensions of our work.

\section{M5-brane preliminaries} 

Our starting point\footnote{We use a mostly plus signature convention 
for the metric and the 3-form fields are dimensionless. We also use hats 
to distinguish the $D=6$ fields and indices from the $D=5$ fields and 
indices.} will be the following six-dimensional symmetric tensor
defined on a single (abelian) M5-brane ($\hat{a},\hat{b} \in (0,1,\ldots,5)$), 
\begin{equation}
\hat{C}^{\hat{a}\hat{b}} = {1 \over K} \, \left[ (1+{1 \over 12} 
\hat{{\cal H}}^2) \, \hat{g}^{\hat{a}\hat{b}}-\, {1\over 4} 
(\hat{{\cal H}}^2)^{\hat{a}\hat{b}} \right] \, ,
\label{bouillat}
\end{equation}
which depends on the gauge invariant 3-form field strength 
$\hat{\cal H} = d\hat{B} + \hat{C}$, where $\hat{B}$ is a 2-form 
gauge field living on the M5-brane and $\hat{C}$ is the 3-form
gauge field of $D=11$ supergravity. 
We also used $(\hat{{\cal H}}^2)^{\hat{a}\hat{b}}\equiv 
\hat{\cal H}^{\hat{a}\hat{c}\hat{d}} 
\left.\hat{\cal H}^{\hat{b}}\right._{\hat{c}\hat{d}}$
and introduced a function $K$ equal to 
\begin{equation}
K=\sqrt{1+{1\over 24} \hat{{\cal H}}^2} \, . 
\label{k}
\end{equation}
This tensor is conformal to the open membrane co-metric\footnote{Because
indices are lowered and raised with the usual metric $\hat{g}_{\hat{a}
\hat{b}}$, we have to make a clear distinction between inverse open 
brane metrics and co-metrics, i.e. $(\hat{G}^{-1})_{\hat{a}\hat{b}} \neq 
\hat{G}_{\hat{a}\hat{b}}$.} and was shown to reduce to the so-called 
Boillat metric of nonlinear DBI electrodynamics in \cite{Gibbons-West}. 

The equations of motion of the tensor multiplet, the nonlinear self-duality 
equation and the energy momentum tensor
on the M5-brane can all be conveniently written using this tensor 
(\ref{bouillat}) \cite{Gibbons-West} (which implies that $C^{-1}$ defines 
the propagation cone for all perturbative degrees of freedom). 
For our purposes we only need 
the nonlinear self-duality equation the 3-form $\hat{{\cal H}}$ satisfies 
on the M5-brane, which is
\begin{equation}
\left. \hat{C}_{\hat{a}} \right.^{\hat{d}} \hat{{\cal H}}_{\hat{d}
\hat{b}\hat{c}} = {\sqrt{-{\rm det}\hat{g}} \over 3!} \epsilon_{\hat{a}
\hat{b}\hat{c}\hat{d}\hat{e}\hat{f}}\hat{{\cal H}}^{\hat{d}\hat{e}\hat{f}} \, .
\label{nlsdeq}
\end{equation} 

For ease of computation we will from now on assume that the M5-brane 
worldvolume is flat, so $\hat{g}^{\hat{a}\hat{b}}=\hat{\eta}^{\hat{a}\hat{b}}$.
In \cite{Howe1, Howe2} many of the calculations made use of 
relating the 3-form $\hat{\cal H}$ to another (unphysical) 
3-form $\hat{h}$ which satisfies
a linear self-duality equation. If possible, we would like to avoid using
the linearly self-dual 3-form $\hat{h}$, but we are going to use it to deduce 
a constraint on $(\hat{\cal H}^4)^{\hat{a}\hat{b}}\equiv (\hat{\cal H}^2)^{
\hat{a}\hat{c}} \left.(\hat{\cal H}^2)^{\hat{b}}\right._{\hat{c}}$, which 
will turn out to be useful.

The two 3-forms are related in the following way
\begin{equation}
\hat{h}_{\hat{a}\hat{b}\hat{c}}= {1\over4}\left.\hat{m}_{\hat{a}}
\right.^{\hat{d}} \hat{\cal H}_{\hat{d}\hat{b}\hat{c}} \, ,
\label{h=mH}
\end{equation}
with
\begin{eqnarray}
\hat{m}^{\hat{a}\hat{b}} &\equiv& \hat{\eta}^{\hat{a}\hat{b}} - 2 
\hat{k}^{\hat{a}\hat{b}} \nonumber \\
\hat{k}^{\hat{a}\hat{b}} &\equiv& \left.\hat{h}^{\hat{a}}\right._{\hat{c}
\hat{d}}\hat{h}^{\hat{b}\hat{c}\hat{d}} \, .
\label{mk}
\end{eqnarray}
Because $\hat{h}$ satisfies a linear self-duality equation we find
first of all that ${\rm Tr}\, \hat{k}^{\hat{a}\hat{b}} \equiv \hat{\eta}
_{\hat{a}\hat{b}} \hat{k}^{\hat{a}\hat{b}} = 0$ and we also deduce
\begin{eqnarray}
\hat{h}_{\hat{a}\hat{b}\hat{e}} \hat{h}^{\hat{c}\hat{d}\hat{e}} &=& 
{1\over 4} \hat{\delta}^{[\hat{c}}_{[\hat{a}} \, \hat{k}^{\hat{d}]}
_{\hat{b}]} \label{hconstraint} \\
\left.\hat{k}_{\hat{a}}\right.^{\hat{c}} \left.\hat{k}_{\hat{c}}\right.
^{\hat{b}} &=& {1\over 6} \hat{k}^2 \, \hat{\delta}_{\hat{a}}^{\hat{b}}\, . 
\label{kconstraint} 
\end{eqnarray}
Using (\ref{kconstraint}) it is straightforward to calculate the 
inverse of $\hat{m}$, which is equal to 
\begin{equation}
(\hat{m}^{-1})^{\hat{a}\hat{b}}= {1\over {1-{2 \over 3}\hat{k}^2}}
(\hat{\eta}^{\hat{a}\hat{b}}+2 \hat{k}^{\hat{a}\hat{b}}) \, .
\label{minverse}
\end{equation}

To find a constraint on $(\hat{\cal H}^4)^{\hat{a}\hat{b}}$ we replace
$\hat{h}$ by $\hat{\cal H}$ in (\ref{hconstraint}) giving us
\begin{equation}
\hat{\cal H}_{\hat{a}\hat{b}\hat{e}} \hat{\cal H}^{\hat{c}\hat{d}\hat{e}} = 
4 {{{2\over3} \hat{k}^2} \over {(1-{2\over3} \hat{k}^2)^2}}  
\hat{\delta}^{[\hat{c}}_{[\hat{a}} \, \hat{\delta}^{\hat{d}]}_{\hat{b}]}
+{4 \over {(1-{2\over3} \hat{k}^2)^2}} \hat{k}^{[\hat{c}}_{[\hat{a}} \, 
\hat{k}^{\hat{d}]}_{\hat{b}]}
+4 {{1+{2\over3} \hat{k}^2} \over {(1-{2\over3} \hat{k}^2)^2}}  
\hat{\delta}^{[\hat{c}}_{[\hat{a}} \, \hat{k}^{\hat{d}]}
_{\hat{b}]} \label{Hconstraint} \, .
\end{equation}
Tracing this equation once allows us to express $\hat{k}^{\hat{a}\hat{b}}$
as
\begin{equation}
\hat{k}^{\hat{a}\hat{b}}= {1\over16} {{(1-{2\over3}\hat{k}^2)^2} 
\over{1+{2\over3} \hat{k}^2}} \left[ (\hat{\cal H}^2)^{\hat{a}\hat{b}} 
- {1\over6} \hat{\cal H}^2 \, \hat{\eta}^{\hat{a}\hat{b}} \right] 
\label{kinH}
\end{equation}
and tracing again gives
\begin{equation}
\hat{\cal H}^2 = 96 \left( {{{2\over 3}\hat{k}^2} \over {(1-{2\over3}\hat{k}^2)^2}} \right) \, ,
\label{k2inH2}
\end{equation}
enabling us to write the right hand side of (\ref{kinH}) solely in terms of
$\hat{\cal H}$. 

So far we have just repeated part of the analysis performed in \cite{Howe1}
and \cite{Gibbons-West}.
To continue we use the expression (\ref{kinH}) and plug that into
equation (\ref{kconstraint}) to find the following expression 
for $(\hat{\cal H}^4)^{\hat{a}\hat{b}}$
\begin{equation}
(\hat{\cal H}^4)^{\hat{a}\hat{b}} = {2\over3} \hat{\cal H}^2 \left[ 
\hat{\eta}^{\hat{a}\hat{b}} + {1\over2} (\hat{\cal H}^2)^{\hat{a}\hat{b}}
\right] \, .
\label{H4constraint}
\end{equation} 
Tracing equation (\ref{H4constraint}) we 
obtain\footnote{Note that we will write $(\hat{\cal H}^2)^2$ to distinguish 
it from $\hat{\cal H}^4$, where the first expression should be understood in 
matrix notation as $\left({\rm Tr}\, (\hat{\cal H}^2)^{\hat{a}\hat{b}}
\right)^2$ and the second as ${\rm Tr}\, (\hat{\cal H}^4)^{\hat{a}\hat{b}}$.} 
\begin{equation}
{1\over4} \hat{\cal H}^4 = \hat{\cal H}^2 (1+{1\over12}\hat{\cal H}^2)\, .
\label{trH4}
\end{equation}
It should be possible to deduce this constraint from the nonlinear 
self-duality equation (\ref{nlsdeq}) directly but we expect that to be 
more elaborate. From now on we will no longer use the 
(unphysical) field $\hat{h}$ (and the tensors that depend on it). 

Using (\ref{H4constraint}) one can easily verify that the 
inverse of $\hat{C}^{\hat{a}\hat{b}}$ is given by
\begin{equation}
(\hat{C}^{-1})_{\hat{a}\hat{b}} = {1 \over K} \, \left[ 
\hat{\eta}_{\hat{a}\hat{b}} + \, {1\over 4} 
(\hat{{\cal H}}^2)_{\hat{a}\hat{b}} \right] \, .
\label{inverseb}
\end{equation}
We note that the traces of $\hat{C}$ and its inverse are both
equal to $6\sqrt{1+{1\over24}\hat{\cal H}^2}$ and that we have, according to 
\cite{Gibbons-West}, the remarkable identity ${\rm det}\, (C^{-1})_{ab} = 
{\rm det}\, g_{ab}$.

We define the open membrane co-metric as
\begin{equation}
\hat{G}^{\hat{a}\hat{b}}_{OM} = z \, \hat{C}^{\hat{a}\hat{b}} 
\label{defOM}
\end{equation}
and it will be our goal in the next section to determine the conformal
factor $z$ by performing the double dimensional reduction.

\section{The open membrane metric on the circle}

First of all we split the $D=6$ indices into $\hat{a}=(a,y)$
where $x^y$ is a compact direction in the worldvolume of the M5-brane 
and identify the $D=5$ (dimensionless) 2-form and 3-form fields as follows
\begin{eqnarray}
\hat{\cal H}_{aby} &\equiv& {\cal F}_{ab} \nonumber \\
\hat{\cal H}_{abc} &\equiv& {\cal H}_{abc} 
\label{redfields}
\end{eqnarray}

As a consequence of the nonlinear self-duality equation in $D=6$ 
(\ref{nlsdeq}) the 3-form ${\cal H}$ and the 2-form ${\cal F}$ 
are related through a set of nonlinear duality equations given by
\begin{eqnarray}
\left. C_a \right.^d {\cal H}_{dbc} + \left. C_a \right.^y {\cal F}_{bc}
&=& {1\over2} \epsilon_{abcde} {\cal F}^{de} \label{5dnld1} \\
\left. C_y \right. ^d {\cal H}_{dab} + \left. C_y \right.^y {\cal F}_{ab}
&=& -{1\over 3!} \epsilon_{abdef} {\cal H}^{def} \label{5dnld2} \\
\left.C_a\right.^d {\cal F}_{db} &=& -{1\over 3!} \epsilon_{abdef} 
{\cal H}^{def} \label{5dnld3} \, , 
\end{eqnarray}
where the different components of $\hat{C}$ are 
\begin{eqnarray}
C^{ab} &\equiv& \hat{C}^{ab} = {1 \over K} \, \left[ (1+{1 \over 12} 
{\cal H}^2 + {1\over4} {\cal F}^2) \, \eta^{ab}-\, {1\over 4} 
({\cal H}^2)^{ab} -{1\over2} ({\cal F}^2)^{ab} \right] 
\label{Cab} \\
C^{ay} &=& {-1\over 4K} \left.{\cal H}^a\right._{cd} {\cal F}^{cd}
\label{Cay} \\
C^{yy} &=& {1\over K} \left( 1+ {1\over 12} {\cal H}^2 \right) \, \eta^{yy} 
\label{Cyy} \, .
\end{eqnarray}
In these expressions we used that $\hat{\cal H}^2 = {\cal H}^2 + 3 
{\cal F}^2$ and $(\hat{\cal H}^2)^{ab} = ({\cal H}^2)^{ab} + 2 
({\cal F}^2)^{ab}$ \footnote{We note that our definition of $({\cal F}^2)^{ab}$
differs by a minus sign with conventional matrix multiplication.}.

Our goal is to express $C^{ab}$ and $C^{yy}$ solely in terms of ${\cal F}$,
using the set of nonlinear duality 
equations (\ref{5dnld1}, \ref{5dnld2}, \ref{5dnld3}). 
Looking at these equations it is 
clear that this is not an easy problem. Instead of solving the equations
(\ref{5dnld1}, \ref{5dnld2}, \ref{5dnld3}) one could also    
look at the reduced expression for $(\hat{\cal H}^4)^{\hat{a}\hat{b}}$ 
(\ref{H4constraint}), giving us the following set of equations
\begin{eqnarray}
\left( \left.({\cal H}^2)^a\right._c + 2 \left.({\cal F}^2)^a\right._c \right)
\left( ({\cal H}^2)^{cb} + 2 ({\cal F}^2)^{cb} \right) 
&+& \left( {\cal H}^{acd} {\cal F}_{cd} \right)  \left( {\cal H}^{bkl} 
{\cal F}_{kl} \right)  \label{1reducH4}\\
= {2\over3}({\cal H}^2 &+& 3 {\cal F}^2)
\left( \eta^{ab} + ({\cal F}^2)^{ab} + {1\over 2} ({\cal H}^2)^{ab} \right) 
\nonumber \\
\left[ {1\over3} {\cal H}^2 \eta^{ac} - \left( ({\cal H}^2)^{ac} + 
2 ({\cal F}^2)^{ac} \right) \right] {\cal H}_{cmn} {\cal F}^{mn} &=& 0 
\label{2reducH4} \\
({\cal H}^{ckl} {\cal F}_{kl})({\cal H}_{cmn} {\cal F}^{mn}) 
- {2\over3}({\cal H}^2 + 3{\cal F}^2) - {1\over3} {\cal H}^2 {\cal F}^2 
&=& 0 
\label{3reducH4}
\end{eqnarray}

The solution to these equations can be found in the most general case 
\cite{Howe2}, but in this paper we will find it useful to simplify 
matters considerably by insisting that the off-diagonal terms in the 
compact direction of the M5-brane Boillat co-metric vanish, i.e.
\begin{equation}
V_c \equiv {\cal H}_{cmn} {\cal F}^{mn} =0 \, . 
\label{HF=0}
\end{equation}
This simplifies the equations considerably and we will proceed by 
focusing our attention on the duality equation (\ref{5dnld2}), where the 
term containing $\left.C_{y}\right.^{d}$ now vanishes. 
Multiplying that equation with $\left.C^{y}\right._{y} {\cal F}^{ac}$
one finds the following useful relations (where the second one is
obtained by tracing the first one\footnote{Actually, using (\ref{HF=0}),  
the second equation can also be deduced directly from (\ref{3reducH4}).})
\begin{eqnarray}
({\cal H}^2)^{ab} &=& {1\over {1+{1\over2}{\cal F}^2}} \left( 
2({\cal F}^2)^{ab} - {\cal F}^2 \eta^{ab} \right) \label{H2F1} \\
{\cal H}^2 &=& { {-3 {\cal F}^2} \over {1+{1\over2}{\cal F}^2}}
\label{H2F2} \, .
\end{eqnarray}
This is all we need to write $C^{ab}$ and $C^{yy}$ strictly in terms 
of ${\cal F}$. As it turns out, we can also find a constraint on 
$({\cal F}^4)^{ab}$. Using both results (\ref{H2F1}, \ref{H2F2}) 
and plugging them into equation (\ref{1reducH4}) we find a 
surprisingly simple expression for $({\cal F}^4)^{ab}$
\begin{equation}
({\cal F}^4)^{ab} = {1\over2} {\cal F}^2 ({\cal F}^2)^{ab} \, .
\label{F4}
\end{equation}
An analysis of this equation in terms of the eigenvalues of ${\cal F}^{ab}$  
quickly reveals that our solutions for ${\cal F}^{ab}$ are 
restricted to rank 2 only. This is a direct consequence of the  
constraint (\ref{HF=0}) we imposed in order to simplify
our analysis. 

We are now ready to express $C^{ab}$ and $C^{yy}$ in terms of 
${\cal F}$ only. We find
\begin{eqnarray}
C^{ab} &=& {1\over \sqrt{1+{1\over2}{\cal F}^2}} \left( (1+{1\over2}{\cal F}^2)
\, \eta^{ab} - ({\cal F}^2)^{ab} \right)
\label{CabF} \\
C^{yy} &=& {\eta^{yy} \over \sqrt{1+{1\over2}{\cal F}^2}} 
\label{CyyF}
\end{eqnarray}
and the inverse of $C^{ab}$, using (\ref{F4}),
\begin{equation}
(C^{-1})_{ab} = {1\over \sqrt{1+{1\over2}{\cal F}^2}} \left(
\eta_{ab} + ({\cal F}^2)_{ab} \right) 
\label{CabFinv} 
\end{equation}
We note that by imposing that the off-diagonal components $C^{ay}$  
vanish (\ref{HF=0}), we would have found
the same result by reducing the inverse M5-brane Boillat metric (instead
of reducing the M5-brane Boillat co-metric). This is no longer true 
when we would not have imposed this constraint.

The symmetric tensor $(C^{-1})_{ab}$ should be equal to the Boillat metric.
At first sight it looks like there is a discrepancy with the result
obtained in \cite{Gibbons-West} where the conformal factor is equal 
to the inverse of $\sqrt{-{\rm det}(\eta_{ab}+{\cal F}_{ab})}$ instead  
of $\sqrt{1+{1\over2}{\cal F}^2}$. However, the 5-dimensional
determinant $-{\rm det}(\eta_{ab}+{\cal F}_{ab})$ can be worked out 
to be equal to 
\begin{equation}
-{\rm det}(\eta_{ab} + {\cal F}_{ab}) = 1+{1\over2}{\cal F}^2 + 
{1\over8} ({\cal F}^2)^2 - {1\over4} {\cal F}^4 \, . 
\label{determinant}
\end{equation}
Tracing the equation (\ref{F4}) it is clear that the last two 
terms in this expression of the determinant cancel each other 
and the right-hand side of (\ref{determinant}) reduces to 
$1+{1\over2}{\cal F}^2$. 
So we conclude that we end up with the expected result, that agrees with 
\cite{Gibbons-West, Howe1}, using a procedure in which we have restricted 
ourselves to consider rank 2 ${\cal F}$ solutions only by imposing 
the (consistent) constraint $V_c = 0$ (\ref{HF=0}).

Our next goal is to determine the conformal factor $z$ (\ref{defOM}).
As discussed in \cite{Gopakumar-OM, Bergshoeff-OM}, the relation
between the 6-dimensional OM-theory parameter $\ell_g$ and the
5-dimensional NCOS parameters $\alpha_{os}, \, \lambda_{os}$, after the
decoupling limit, mimics the bulk relations between M-theory 
and IIA string theory. An important assumption we will make is that these 
relations continue to hold beyond the decoupling limit. 
As a result, in analogy with the bulk relation 
$\eta^{yy}\equiv{g_s}^{-{4\over3}}$, we will define the open string coupling 
constant $\lambda_{os}$ as follows 
\begin{equation}
\lambda_{os}^{-{4\over3}} \equiv G^{yy}_{OM} = z\, C^{yy} = 
{ {z\,g_s^{-{4\over3}}} \over \sqrt{{1+{1\over2}{\cal F}^2}}} \, . 
\label{defGos}
\end{equation}
We can now fix $z$ by demanding this expression to be
equal to the Seiberg-Witten one \cite{Seiberg-Witten} (replacing
the determinant with our result $1+{1\over 2}{\cal F}^2$) 
\begin{equation}
\lambda_{os} = g_s \sqrt{{1+{1\over2}{\cal F}^2}} \, .
\label{SWGos}
\end{equation}
This determines $z$ to be equal to 
\begin{equation}
z= ({1+{1\over2}{\cal F}^2})^{-{1\over6}} \, .
\label{z}
\end{equation}

The next thing we should show is that upon 
including $g_s$'s everywhere in our expressions for $C^{ab}$ we precisely find
the open string metric and the open string coupling upon reduction
of the open membrane metric.
First of all we define the open membrane metric and the
open string metric to be related in the following way (analogous to 
the well-known bulk relation $\eta_{ab}=g_s^{-2/3} \eta_{ab}^{(s)}$)
\begin{equation}
(G^{-1}_{OM})_{ab} \equiv z^{-1} (C^{-1})_{ab} \equiv 
\lambda_{os}^{-{2\over3}} \, (G^{-1}_{os})_{ab} \, .
\label{osmetricdef}
\end{equation}
Using this definition, plugging in our expressions for $z$ and 
$\lambda_{os}$ we indeed find precisely the open string metric
\begin{equation}
(G^{-1}_{os})_{ab} = \eta_{ab}^{(s)} + ({\cal F}^2)_{ab} \, ,
\label{osmetric}
\end{equation}
where $({\cal F}^2)_{ab}$ is now defined
with respect to the string frame metric, i.e. $({\cal F}^2)_{ab} = 
{\cal F}_{ac} \, \eta^{cd}_{(s)} \, {\cal F}_{bd}$. We have now 
successfully shown that the open membrane metric, with  
the previously unknown conformal factor $z$ now determined, 
reduces to the Seiberg-Witten expressions for the open string metric and the 
open string coupling. We note that our procedure that lead to this result 
was restricted to rank 2 ${\cal F}$ only and assumed reduction ansatze 
analogous to the bulk.

The only thing left to do is to rewrite this conformal factor $z$
(\ref{z}) in terms of $\hat{\cal H}$, using (\ref{H2F2}) and the fact 
that $\hat{\cal H}^2={\cal H}^2 + 3 {\cal F}^2$. It turns out to
be useful to write this expression in terms of $K$ (\ref{k}) and we
find
\begin{equation}
z= \left( (2K^2 -1) \pm 2K^2 \sqrt{1-K^{-2}} \right)^{-{1\over6}} \, .
\label{zinK}
\end{equation}
Because the relation between $\hat{\cal H}^2$ and ${\cal F}^2$ is
quadratic there is a sign ambiguity in this expression. 

The sign ambiguity can be fixed by performing the following 
consistency check. Our result for the conformal factor
should reproduce the result reported
in \cite{Bergshoeff-OM} that the conformal factor scales as
$\left({{\ell_p} \over {\ell_g}}\right)^2$ in the OM-theory decoupling limit.
One can check that in the OM-theory limit $K^2$ scales as
\begin{equation}
K^2 = 1+{1\over 24} \hat{\cal H}^2 \sim \left({{\ell_g} \over 
{\ell_p}}\right)^3 \, .
\label{Kscale}
\end{equation}
As it turns out, only when we choose the $(-)$ sign in the expression for
$z$ (\ref{zinK}), due to some crucial cancellations (expanding the square 
root to second order), the conformal factor precisely scales as
\begin{equation}
{\left[ (2K^2 -1) - 2K^2 \sqrt{1-K^{-2}} \right]^{1\over6} \over K}
\sim (K^2)^{-{2\over3}} = \left({{\ell_p} \over {\ell_g}}\right)^2 \, ,
\label{confscale}
\end{equation}
therefore reproducing the result of \cite{Bergshoeff-OM}. 
Taking the $(+)$ sign
however, results in a diverging open membrane metric with respect to
the Planck length $\ell_p$ (the conformal factor is proportional to 
${\ell_p \over \ell_g}$ to leading order) in the OM-theory decoupling limit. 
So, finally, after getting rid of the sign ambiguity to agree with
the expected behavior in the decoupling limit, the open membrane
metric is determined to be equal to
\begin{equation}
(G^{-1}_{OM})_{ab} = {\left( (2K^2 -1) - 2K^2 \sqrt{1-K^{-2}} 
\right)^{1\over6} \over K} \, \left[ \hat{\eta}_{\hat{a}\hat{b}} 
+ \, {1\over 4} (\hat{{\cal H}}^2)_{\hat{a}\hat{b}} \right] \, .
\label{OMexpr}
\end{equation}

\section{The open membrane metric on the torus}

Our results support the idea that the open membrane metric 
can be understood, upon reduction, as providing the geometric origin 
of the open string (or more generally, open brane) moduli, 
independent of whether the M5-brane or D-brane is decoupled from the bulk 
or not. This also implies the existence of an 
$SL(2,Z)$ generalization of the open string coupling constant, i.e.
a complex open string coupling constant, which should equal the
modular parameter of the (OM) torus, analogous to what
happens when wrapping M-theory on a torus. 
  
We will consider $z^i$, with $i=4,5$, to 
be the coordinates on the torus and $x^{a}$, with $a=0,1,\ldots, 3$ as
the (directly reduced) D3-brane coordinates. From the start we will assume that
the off-diagonal open membrane (co-) metric components $\hat{G}^{ai}$ 
vanish, again effectively restricting us to rank 2 solutions. For our 
purposes here we will also assume that after 
reduction on the $T^2$ we will be left with 2-forms 
only\footnote{One can check that after imposing $\hat{G}_{ai}=0$ 
the duality equations relating the 2-forms `decouple' from the duality 
equations relating the 1- and 3-forms, allowing for a consistent 
truncation with vanishing 1- and 3-forms.}. So we define 
\begin{eqnarray}
\hat{\cal H}^{abi} &\equiv& {\cal F}^{ab, \, i} \label{2BdefF} \\
\hat{\cal H}^{aij} &\equiv& {\cal V}^a \equiv 0 \nonumber \\
\hat{\cal H}^{abc} &\equiv& {\cal H}^{abc} \equiv 0 \nonumber \, . 
\end{eqnarray}
Because we restricted the fields ${\cal H}$ and ${\cal V}$ to be 
zero, we can write 
\begin{equation}
(\hat{\cal H}^2)^{ab}= 2\, ({\cal F}^{i} \zeta_{ij} {\cal F}^j)^{ab} 
\label{2BHF}
\end{equation} 
and
\begin{equation}
\hat{\cal H}^2 = 3\, {\cal F}^{i} \zeta_{ij} {\cal F}^{j} \, ,
\label{2BHF2}
\end{equation}
where $\zeta_{ij}$ is the metric on the torus (it is understood
that the absence of $D=4$ spacetime indices means they are summed over). 
Obviously these
expressions are $SL(2,R)$ invariant. Again, as a consequence of
the nonlinear self-duality equation on the M5-brane, the doublet
of two-forms on the D3-brane has to satisfy the following set 
of (self-) duality equations
\begin{eqnarray}
C^{ij} {\cal F}_{ab,\, j} &=& {1\over 2} \epsilon_{abef} (\epsilon^{ij}
\left.{\cal F}^{ef}\right._j) \label{nlsdt2} \\
C^{ad} {\cal F}_{db,\, i} &=& {1\over 2} \epsilon_{abef} (\epsilon_{ij}
\,{\cal F}^{ef,\, j}) \label{nlsd3} \, ,
\end{eqnarray} 
where the tensors $C$ are now given by
\begin{eqnarray}
C^{ij} &=& {1\over K} \left[ (1+{1\over 4} {\cal F}^{k} \zeta_{kl} 
{\cal F}^{l}) \, \zeta^{ij} - {1\over 4} ({\cal F}^2)^{ij} \right]  
\label{Cij} \\
C^{ab} &=& {1\over K} \left[ (1+{1\over 4} {\cal F}^{k} \zeta_{kl} 
{\cal F}^{l})\, \eta^{ab} - {1\over2} ({\cal F}^{k}
\zeta_{kl} {\cal F}^{l})^{ab} \right] \, .
\label{T2Cab}
\end{eqnarray}

The reduction of equation (\ref{H4constraint}) gives the following relations
for $({\cal F}^2)^{ik} \zeta_{kl} ({\cal F}^2)^{jk}$ and
$\left(({\cal F}^k \zeta_{kl} {\cal F}^{l})^2\right)^{ab}$ 
\begin{eqnarray}
({\cal F}^2)^{ik} \zeta_{kl} ({\cal F}^2)^{jl} &=& 2 ({\cal F}^k \zeta_{kl} 
{\cal F}^l) \left[ \zeta^{ij} + {1\over2} ({\cal F}^2)^{ij} \right] 
\label{F4ij} \\
\left(({\cal F}^k \zeta_{kl} {\cal F}^{l})^2\right)^{ab} &=& {1\over2}
({\cal F}^k \zeta_{kl} {\cal F}^l) \left[ \eta^{ab} + ({\cal F}^k \zeta_{kl} 
{\cal F}^l)^{ab} \right] \label{T2F4ab} \, ,
\end{eqnarray}
which are useful for checking that the inverses of $C$ on the 
torus and the D3-brane are given by
\begin{eqnarray}
(C^{-1})_{ij} &=& {1\over K} \left[ \zeta_{ij} + {1\over4} ({\cal F}^2)_{ij}
\right] \label{inverseCij} \\
(C^{-1})_{ab} &=& {1\over K} \left[ \eta_{ab} + {1\over 2} ({\cal F}^k 
\zeta_{kl} {\cal F}^l)_{ab} \right] \, . \label{T2inverseCab}
\end{eqnarray}

Previously, when reducing the M5-brane on a circle we used the duality
equations to rewrite everything in terms of ${\cal F}$ only and we 
obtained the expected Seiberg-Witten results. However, in this case 
the set of duality equations generically relate the $SL(2,R)$ doublet 
${\cal F}^i$ to the doublet ${\cal F}^i$ in a complicated (intertwined) 
way, i.e. it is more like a self-duality equation. 
Another way of saying this is that generically both electric and magnetic 
components of both field strengths in the doublet have to be turned on to 
satisfy the set of duality equations (\ref{nlsdt2}) and (\ref{nlsd3}). 
The question arises as to how we can obtain the Seiberg-Witten
results for the D3-brane, which should still be valid when  
only one of the 2-forms in the doublet is turned on\footnote{To be more 
precise, we need the off-diagonal components of $C^{ij}$ to vanish. 
As we will see, this means that the bulk axion has to vanish as well.}.  

The answer is that in the special case where the off-diagonal components
of $C^{ij}$ vanish, the set of duality 
equations allows for solutions with only one of the 2-forms turned on in a 
particular direction, i.e. the set of self-duality equations reduces to a 
set of ordinary (nonlinear) duality equations relating one of the 2-forms 
in the doublet to the other one. This 
enables us to follow the same procedure as before, solving for just one
of the 2-forms in the doublet and we will check that 
this indeed gives the expected result for the open string metric and coupling 
constant. Another way to understand this is that in a generic $SL(2,R)$ 
basis we are clearly forced to use both 2-forms and the result for the 
open membrane metric is proportional to (\ref{inverseCij}), whereas in the
special case we should be able to rewrite (\ref{inverseCij}) 
in terms of just one of the 2-forms and it is only then  
that we will find the Seiberg-Witten results. It must be clear that on
the D3-brane it is always possible to find such an $SL(2,R)$ 
basis\footnote{Upon considering $SL(2,Z)$ this is no longer 
necessarily true, in that case one needs a rational axion to be able to 
rotate to a frame in which the axion vanishes.}, at the cost of giving up 
manifest $SL(2,R)$ invariance.

After clearing that up we now want to rewrite our results (\ref{inverseCij})
and (\ref{T2inverseCab}) in terms of the appropriate open string 
quantities, the open string (``Einstein'') metric $G^{-1}_{osE}$ and 
the {\it complex} open string coupling constant $T = X + i\, 
\lambda_{os}^{-1}$. First of all we make the following standard
identifications of the bulk quantities ($\tau = \chi + i\, g_s^{-1}$)
\begin{eqnarray}
\eta_{ab} &=& {1\over \sqrt{A^{\rm (bulk)}_{T^2}}} \, \eta_{ab}^{(E)} 
\label{etaE} \\
A^{\rm (bulk)}_{T^2} &=& \sqrt{{\rm det}\, \zeta_{ij}} = \sqrt{\zeta_{44}
\zeta_{55} - \zeta_{45}^2}
\label{bulkarea} \\
\chi &=& - {\zeta_{45} \over \zeta_{44}} \label{axionb} \\
g_s^{-2} &=& {{\zeta_{55} - {\zeta_{45}^2 \over \zeta_{44}}} 
\over \zeta_{44}} \label{gsdef} \, . 
\end{eqnarray}
Together with (\ref{2BdefF}), this enables us to write the $SL(2,R)$ 
invariant quantity $({\cal F}^i \zeta_{ij} {\cal F}^j)$ as follows
\begin{equation}
({\cal F}^i \zeta_{ij}{\cal F}^j) = g_s ({\cal F}^4 - \chi {\cal F}^5)^2 
+ g_s^{-1} ({\cal F}^5)^2 \, ,
\label{Finvariant}
\end{equation}
which is invariant under the following $SL(2,R)$ transformations
\begin{eqnarray}
\tau &\rightarrow& {{a \tau + b} \over {c \tau +d}} \label{sltau} \\
\left( 	\begin{array}{c} 
	{\cal F}^4 \\
	{\cal F}^5 
	\end{array} \right) &\rightarrow& \,  
\left(	\begin{array}{cc}
	\,a\, & \,b\, \\
	\,c\, & \,d\, 
	\end{array} \right) \,
\left( 	\begin{array}{c} 
	{\cal F}^4 \\
	{\cal F}^5 
	\end{array} \right) \label{slvector} \, ,
\end{eqnarray}
with $ad-bc=1$.
We note that when calculating $({\cal F}^i \zeta_{ij} {\cal F}^j)_{ab}$, 
because we defined the two-forms contravariantly (\ref{2BdefF}), 
we obtain an extra factor $(A_{T^2}^{(bulk)})^{-{1\over2}}$ as compared to 
equation (\ref{Finvariant}).

The next step is to define the corresponding open string quantities
on the D3-brane in an analogous way using the open membrane metric
(\ref{OMexpr}). So we define
\begin{eqnarray}
(G^{-1}_{OM})_{ab} &\equiv& {1\over \sqrt{A^{\rm (OM)}_{T^2}}} \, 
(G^{-1}_{osE})_{ab} 
\label{osmetricE} \\
A^{\rm (OM)}_{T^2} &=& \sqrt{{\rm det}\, (G^{-1}_{OM})_{ij}} = 
\sqrt{(G^{-1}_{OM})_{44} (G^{-1}_{OM})_{55} - (G^{-1}_{OM})_{45}^2} 
\label{OMarea} \\
X &\equiv& - {(G^{-1}_{OM})_{45} \over (G^{-1}_{OM})_{44}} 
\label{axionOM} \\
\lambda_{os}^{-2} &\equiv& {{(G^{-1}_{OM})_{55} - 
{(G^{-1}_{OM})_{45}^2 \over (G^{-1}_{OM})_{44}}} 
\over (G^{-1}_{OM})_{44}} \label{gosdefT2} \, . 
\end{eqnarray}

Using these definitions and (\ref{2BdefF}), we find the following
results for the D3-brane open string quantities expressed in terms
of the appropriate bulk quantities (\ref{etaE})-(\ref{gsdef})\footnote{We
apologize for using the numbers $4$ and $5$ to denote the different 
components of the $SL(2,R)$ vector, which should not be confused with taking
fourth or fifth powers of ${\cal F}$.}
\begin{eqnarray}
(G^{-1}_{osE})_{ab} &=& {1\over z\,K} \sqrt{{A^{\rm (OM)}_{T^2}} \over
{A^{\rm (bulk)}_{T^2}}} \left[ \eta_{ab}^{(E)} + {1\over 2} \left( g_s
({\cal F}^4 - \chi {\cal F}^5)^2 + g_s^{-1} ({\cal F}^5)^2 
\right)_{ab}\right] \label{osemetric} \\
X &=& {{\chi +{1\over 4} g_s (\chi {\cal F}^4 - |\tau|^2 {\cal F}^5)_{ab} 
({\cal F}^4 - \chi {\cal F}^5)^{ab}} 
\over {1+ {1\over 4} g_s ({\cal F}^4 - \chi {\cal F}^5 )^2}}
\label{osaxion} \\
\lambda_{os}^{-2} &=& {{|\tau|^2 + {1\over 4} g_s  
( \chi {\cal F}^4 - |\tau|^2 {\cal F}^5 )^2} \over 
{1+ {1\over 4} g_s ({\cal F}^4 - \chi {\cal F}^5 )^2}}
- X^2 \label{oscoupling} \, ,
\end{eqnarray}
where it should be understood that all contractions of $D=4$ indices
are now taken with respect to the Einstein frame metric $\eta_{ab}^{(E)}$
and where $z$ is the conformal factor that was determined in the previous 
section (\ref{zinK}).
For completeness we should also give the expressions for $K^2$ (\ref{k}) 
and $A^{\rm (OM)}_{T^2}$ (in terms of bulk quantities)
\begin{eqnarray}
K^2 &=& 1+{1\over 8} \left( g_s ({\cal F}^4 - \chi {\cal F}^5 )^2 + g_s^{-1} 
({\cal F}^5)^2 \right) \label{KinIIB} \\
A^{\rm (OM)}_{T^2} &=& (G^{-1}_{OM})_{44} \, \lambda_{os}^{-1} = 
{{A^{\rm (bulk)}_{T^2} \, g_s} \over {z\, K}}
\left( 1+ {1\over 4} g_s ({\cal F}^4 - \chi {\cal F}^5 )^2 \right) 
\, \lambda_{os}^{-1} 
\label{OMareaT2} \, .
\end{eqnarray}

As a first check that these expressions are correct we observe that
S-duality transformations of bulk quantities induce S-duality 
transformations on the D3-brane. One can check that upon the 
S-duality transformation
\begin{equation}
{\cal F}^4 \leftrightarrow {\cal F}^5 \, , \quad \tau \rightarrow 
{1\over \tau}  \, , 
\label{bulkS}
\end{equation}
one induces the following S-duality transformation on the open string
modular parameter $T$  
\begin{equation}
T \rightarrow {1\over T} \, . 
\label{braneS}
\end{equation}
Another straightforward check is to consider shift transformations, i.e. 
\begin{equation}
{\cal F}^4 \rightarrow {\cal F}^4 + b {\cal F}^5 \, , \quad {\cal F}^5 
\rightarrow {\cal F}^5 \, , \quad \tau \rightarrow \tau + b \, .  
\label{bulk+}
\end{equation}
Although perhaps not immediately obvious these transformations 
indeed induce shift transformations on the open string modular 
parameter $T$, i.e.
\begin{equation}
T \rightarrow T + b \, .  
\label{brane+}
\end{equation}
Based on these explicit checks we are therefore confident that the full 
group of $SL(2,R)$ transformations is indeed induced from the closed
string modular parameter $\tau$ onto the open string modular 
parameter $T$ (also transforming the $SL(2,R)$ vector ${\cal F}^i$), 
as it should.

We now want to check 
that when we take $\chi=0$ and $({\cal F}^4)_{ab} ({\cal F}^5)^{ab}=0$ 
(giving $X=0$ and necessarily giving up manifest $SL(2,R)$ invariance) 
we obtain the Seiberg-Witten results. 
This should involve solving ${\cal F}^4$ in terms of ${\cal F}^5$ 
with the help of the duality equations (\ref{nlsdt2}) (solving for 
${\cal F}^5$ we expect to find the S-dual result). 
Setting $\chi=0$ in 
(\ref{osemetric})-(\ref{oscoupling}) gives us
\begin{eqnarray}
(G^{-1}_{osE})_{ab} &=& {1\over z\,K} \sqrt{{A^{\rm (OM)}_{T^2}} \over
{A^{\rm (bulk)}_{T^2}}} \left[ \eta_{ab}^{(E)} + {1\over 2}
\left( g_s ({\cal F}^4)^2 + g_s^{-1} ({\cal F}^5)^2 \right)_{ab} \right] 
\label{x=0metric} \\
\lambda_{os}^{-2} &=& g_s^{-2} \left( {{1 + {1\over 4} g_s^{-1} 
({\cal F}^5)^2} \over {1+ {1\over 4} g_s ({\cal F}^4)^2}} \right)
\label{x=0coupling} \, .
\end{eqnarray}
When $X=0$ (giving us only diagonal entries in $C^{ij}$) we can 
now use the duality equation (\ref{nlsdt2}) 
and/or equation (\ref{F4ij}) to write ${\cal F}^5$ in terms of 
${\cal F}^4$. The appropriate relations that can be deduced are
\begin{eqnarray} 
g_s^{-1} ({\cal F}^5)^2_{ab} &=& {1 \over {1+{1\over 2}
g_s ({\cal F}^4)^2}} \left[ g_s ({\cal F}^4)^2_{ab} -
{1\over 2} g_s ({\cal F}^4)^2 \eta_{ab}^{(E)} \right] \label{x=0F4F5ab} \\
g_s^{-1} ({\cal F}^5)^2 &=& {{- g_s ({\cal F}^4)^2} \over {1+{1\over 2} 
g_s ({\cal F}^4)^2}} \label{x=0F4F5} \, . 
\end{eqnarray}
We note that by performing an S-duality transformation $g_s ({\cal F}^4)^2 
\rightarrow g_s^{-1} ({\cal F}^5)^2$ we obtain the other set of equations 
(which are of course redundant because they follow uniquely from 
(\ref{x=0F4F5ab}) and (\ref{x=0F4F5})). 
These equations can first of all be used to find the following expression
for the conformal factor in $(G^{-1}_{osE})_{ab}$ (\ref{osemetric}) (also
using (\ref{zinK}), (\ref{KinIIB}) and (\ref{OMareaT2})) 
\begin{equation}
{1\over z\,K} \sqrt{{A^{\rm (OM)}_{T^2}} \over {A^{\rm (bulk)}_{T^2}}} = 
{{\left(1+{1\over 2} g_s ({\cal F}^4)^2 \right)^{3\over4}} \over
{1+{1\over 4} g_s ({\cal F}^4)^2}} \label{cf} \, .
\end{equation}
Plugging these expressions (\ref{x=0F4F5ab})-(\ref{cf}) 
into the D3-brane open string metric 
(\ref{x=0metric}) and open string coupling (\ref{x=0coupling}) we find
the following results
\begin{eqnarray}
(G^{-1}_{osE})_{ab} &=& \left( 1+{1\over 2} g_s ({\cal F}^4)^2 
\right)^{-{1\over4}} \left[ \eta_{ab}^{(E)} + g_s ({\cal F}^4)^2_{ab} 
\right] \label{metricF4} \\
\lambda_{os} &=& g_s \sqrt{1+ {1\over 2} g_s ({\cal F}^4)^2}
\label{couplingF4} \, .
\end{eqnarray}
We should keep in mind that until now we have been using a contravariant
definition of the 2-form field strengths (\ref{2BdefF}), which is not
standard. 
To transform to the standard covariant definition one actually needs 
to include a factor of $g_s^{-2} = (A_{T^2}^{(bulk)})^2 (\zeta^{44})^2$ 
in all $({\cal F}^4)^2$ terms. Doing this and realizing
that we should now transform to the string frame metric,
using $\eta_{ab}^{(s)}=g_s^{1\over2} \eta_{ab}^{(E)}$ and defining 
analogously
\begin{equation}
(G^{-1}_{os})_{ab} \equiv \lambda_{os}^{1\over 2} \, (G^{-1}_{osE})_{ab} \, ,
\label{OSframe}
\end{equation}
we finally obtain the open string metric and coupling, as promised,
\begin{eqnarray}
(G^{-1}_{os})_{ab} &=& \eta_{ab}^{(s)} +  ({\cal F}_4)^2_{ab}
\label{T2xmetric} \\
\lambda_{os} &=& g_s \sqrt{1+ {1\over 2} ({\cal F}_4)^2}
\label{T2xgos} \, .
\end{eqnarray}

As mentioned before, when solving in terms of ${\cal F}^5$ instead, 
we expect to find the S-dual result. We want to emphasize however that 
performing an S-duality transformation is definitely not the same as 
solving in terms of the other field in the doublet. Using the duality 
equations one interchanges electric and 
magnetic components, whereas the S-duality transformations do not 
interchange electric and magnetic components.
The fact that we do find the S-dual result explains why on the D3-brane 
S-duality can be effectively represented by an 
interchange of electric and magnetic field strengths; starting with an 
electric ${\cal F}^4$, S-duality gives us an electric ${\cal F}^5$, 
but by using the duality equations we 
can relate this electric ${\cal F}^5$ to a magnetic ${\cal F}^4$ giving us 
an effective interchange of electric and magnetic components of ${\cal F}^4$ 
upon S-duality. 
Starting from (\ref{x=0metric}) 
and (\ref{x=0coupling}) and now solving for ${\cal F}^5$ we indeed find
\begin{eqnarray}
(G^{-1}_{os})_{ab} &=& {\lambda_{os} \over g_s}
\left[ \eta_{ab}^{(s)} + g_s^2 ({\cal F}_5)^2_{ab} \right]
\label{metricF5} \\
\lambda_{os} &=& {g_s \over \sqrt{1+ {1\over 2} g_s^2 ({\cal F}_5)^2}}
\label{gosF5} \, ,
\end{eqnarray}
which can easily be checked to be equivalent to performing an S-duality 
transformation on (\ref{T2xmetric}) and (\ref{T2xgos}) (remember that
the string frame metric transforms as $\eta_{ab}^{(s)} \rightarrow
{\eta_{ab}^{(s)} \over g_s}$). It seems natural to relate the above 
result to an open D-string metric $(G^{-1}_{od})_{ab}$ and open D-string
coupling, defined in the following way
\begin{equation}
(G^{-1}_{od})_{ab} \equiv {(G^{-1}_{os})_{ab} \over \lambda_{os}} \, , \quad
\lambda_{od} \equiv {1\over \lambda_{os}} \, .
\label{odmod}
\end{equation} 
Also identifying ${\eta^{(s)} \over g_s} \equiv \eta^{(d)}$ and rewriting 
the square of ${\cal F}^5$ with respect to this metric, we can rewrite 
the expressions (\ref{metricF5}) and (\ref{gosF5}) as
\begin{eqnarray}
(G^{-1}_{od})_{ab} &=& \eta_{ab}^{(d)} + ({\cal F}_5)^2_{ab} 
\label{odmetricF5} \\
{1 \over \lambda_{od}} &=& \lambda_{os} = {g_s \over \sqrt{1+ {1\over 2} 
({\cal F}_5)^2}} \label{godF5} \, .
\end{eqnarray}
This metric can also be used to describe the same physics on the D3-brane, 
a priori there is no reason to prefer using the open string metric 
instead of the open D-string metric. Only when considering a particular
limit of the closed string moduli will one create a distinction. 
To give a concrete example; it is not hard to show that 
the NCOS limit in (\ref{T2xmetric}) and (\ref{T2xgos}) maps to the open 
D-string NCYM limit in (\ref{odmetricF5}) and (\ref{godF5}), 
illustrating the fact that the NCOS theory can equivalently be described 
by an open D-string NCYM theory (and vice-versa of course) 
\cite{Lu, Larsson-Sundell}.   

Going back to our $SL(2,R)$ covariant expressions (\ref{osemetric}),
(\ref{osaxion}) and (\ref{oscoupling}) it seems natural to interpret
these as a result of probing the D3-brane with open $(p,q)$ 
strings \cite{Schwarz-pq}.
The existence of an OM-theory decoupling limit should at first sight guarantee
the existence of a correspondingly well-defined decoupling limit 
for $(p,q)$ open strings ending on the D3-brane. Indeed, the 
situation is very similar to the M5-brane, because
generically our doublet of 2-forms satisfies a nonlinear
self-duality equation (\ref{nlsd3}). This means we will have to consider 
constant 2-form doublet field strengths in all directions (electric and 
magnetic) on the D3-brane. It should be possible to deduce a parametrisation 
of these (constant flux) solutions by reducing the M5-brane parametrisation
as given in \cite{Bergshoeff-NCM5}, but we will not do so here \cite{todo}.
Another observation strengthening our belief that the decoupling limit
for $(p,q)$ strings is similar to the one on the M5-brane, is that in the 
OM-theory limit the conformal factor of the $(p,q)$ open string metric 
(\ref{osemetric}) scales as 
\begin{equation}
{1\over z\,K} \sqrt{{A^{\rm (OM)}_{T^2}} \over {A^{\rm (bulk)}_{T^2}}}
\sim \left({l_p \over l_g}\right)^3 \propto {\alpha^\prime \over 
\alpha^\prime_{(eff)}} \label{propscale} \, .
\end{equation}
This means that in order to obtain a finite length scale 
$\alpha^\prime_{(eff)}$ the other part of the open $(p,q)$ string 
metric (\ref{osemetric}) should be fixed in the decoupling limit, 
analogous to what happens 
in the OM-theory limit on the M5-brane.

\section{Summary and discussion}

For the readers convenience let us start this section 
by summarizing the main results presented in this paper. 
To fix the conformal factor of the open membrane metric we first of all 
assumed the following reduction ansatze (in analogy to the bulk) between 
the open membrane and open string moduli
\begin{eqnarray}
(G^{-1}_{OM})_{ab} &\equiv& \lambda_{os}^{-{2\over3}} \, 
(G^{-1}_{os})_{ab} \label {GOMGos} \\
(G^{-1}_{OM})_{yy} &\equiv& \lambda_{os}^{4\over3} 
\label{GyyGos} \, . 
\end{eqnarray}
Upon reduction we effectively restricted our attention to rank 2 solutions 
by imposing the (simplifying) constraint ${\cal H}^{amn}{\cal F}_{mn}=0$.
Using one of the two reduction ansatze fixes the conformal factor of the open 
membrane metric in order to reproduce one of the Seiberg-Witten results, 
the other definition then consistently reproduces the other Seiberg-Witten
result. In the reduction procedure we made 
important use of the following relations, which are consequences 
of the (self-) duality equations and imposing ${\cal H}^{amn}{\cal F}_{mn}=0$
upon reduction
\begin{eqnarray}
(\hat{\cal H}^4)^{\hat{a}\hat{b}} &=& {2\over3} \hat{\cal H}^2 \left[ 
\hat{\eta}^{\hat{a}\hat{b}} + {1\over2} (\hat{\cal H}^2)^{\hat{a}\hat{b}}
\right] \\
({\cal H}^2)^{ab} &=& {1\over {1+{1\over2}{\cal F}^2}} \left( 
2({\cal F}^2)^{ab} - {\cal F}^2 \eta^{ab} \right) \\
({\cal F}^4)^{ab} &=& {1\over2} {\cal F}^2 ({\cal F}^2)^{ab} \, ,
\end{eqnarray}
where the last constraint on $({\cal F}^4)^{ab}$ implies that we have 
restricted ourselves to rank 2 solutions only and 
\begin{equation}
\sqrt{-{\rm det}(\eta_{ab} + {\cal F}_{ab})} = \sqrt{1+{1\over2}{\cal F}^2}
\label{sqrtdet} \, . 
\end{equation}
Using all this information we showed that the following open membrane metric
\begin{equation}
(G^{-1}_{OM})_{ab} = {\left( (2K^2 -1) - 2K^2 \sqrt{1-K^{-2}} 
\right)^{1\over6} \over K} \, \left[ \hat{\eta}_{\hat{a}\hat{b}} 
+ \, {1\over 4} (\hat{{\cal H}}^2)_{\hat{a}\hat{b}} \right] 
\end{equation}
indeed scales as anticipated in
\cite{Bergshoeff-OM} in the OM-theory decoupling limit and reproduces
the open string metric and coupling as first given in \cite{Seiberg-Witten} 
upon double dimensional reduction (for rank 2 solutions).

One thing to worry about is whether our solution for the conformal 
factor $z$ depends on our restriction to rank 2 solutions. 
Because from the point of view of the M5-brane the difference between rank 2 
and rank 4 solutions is nothing but a trivial rotation, we would argue that 
our final result is independent of that restriction. However, it would 
be interesting and worthwhile to show this by actually performing the 
reduction in this more general (and more complicated) case and we hope to 
report on this (and some other issues) in the near future \cite{todo}.   

After fixing the conformal factor we performed a double dimensional 
reduction of the M5-brane on a 2-torus (again effectively restricting 
our attention to rank 2 solutions only) to obtain manifestly $SL(2,R)$ 
covariant results for the open string metric and complex open string 
coupling $T=X+i \lambda_{os}$
\begin{eqnarray}
(G^{-1}_{osE})_{ab} &=& {1\over z\,K} \sqrt{{A^{\rm (OM)}_{T^2}} \over
{A^{\rm (bulk)}_{T^2}}} \left[ \eta_{ab}^{(E)} + {1\over 2} \left( g_s
({\cal F}^4 - \chi {\cal F}^5)^2 + g_s^{-1} ({\cal F}^5)^2 
\right)_{ab}\right]  \\
X &=& {{\chi +{1\over 4} g_s (\chi {\cal F}^4 - |\tau|^2 {\cal F}^5)_{ab} 
({\cal F}^4 - \chi {\cal F}^5)^{ab}} 
\over {1+ {1\over 4} g_s ({\cal F}^4 - \chi {\cal F}^5 )^2}} \\
\lambda_{os}^{-2} &=& {{|\tau|^2 + {1\over 4} g_s  
( \chi {\cal F}^4 - |\tau|^2 {\cal F}^5 )^2} \over 
{1+ {1\over 4} g_s ({\cal F}^4 - \chi {\cal F}^5 )^2}}
- X^2 \, .
\end{eqnarray}
We showed that S-duality and shift transformations of the bulk quantities
induce S-duality and shift transformations on the D3-brane quantities, 
as they should. To reproduce the Seiberg-Witten NCOS metric and 
coupling we used the nonlinear
duality equations on the D3-brane and we emphasized the important 
role played by the duality equations in the context of manifest 
$SL(2,R)$ invariance on the D3-brane. Here we would also like to anticipate 
the existence of an elegant manifestly $SL(2,R)$ covariant formulation 
of the (single) D3-brane equations of motion (and perhaps even an action) 
in terms of the $SL(2,R)$ invariant open string metric. 
The natural interpretation 
of our $SL(2,R)$ covariant results would be in terms of 
(non-commutative) $(p,q)$ open strings and we gave some (mainly OM-theory) 
arguments to suggest that a $(p,q)$ open string decoupling limit should 
exist. 

One thing we did not consider in this paper is the non-commutativity 
parameter $\theta_{ab}$. We concentrated our attention on the open 
string metrics and coupling as they can be directly related to the reduced 
open membrane metric. As explained in \cite{Gibbons-Herdeiro}, there
does exist a nonlinear electrodynamics analogue of the non-commutativity
parameter which is related to the dual Maxwell field. It would be
interesting to find the expression for the non-commutativity parameter
in our $SL(2,R)$ covariant case. Also, it would be very interesting to 
see if and how we can generalize this relation between the non-commutativity
parameter and the dual Maxwell field to the M5-brane and the self-dual
tensor multiplet. In \cite{Gibbons-West} a 3-form $P$ was introduced on the 
M5-brane which could have all the properties we are looking for. 
We hope to report on this interesting possibility in a 
future publication \cite{todo}.

As a consequence of quantum effects one expects the $SL(2,R)$ symmetry
group to reduce to $SL(2,Z)$. One immediate consequence would be that
only when the axion is rational can we $SL(2,Z)$ rotate into a frame in
which the axion vanishes. The same is therefore true for the 
D3-brane (open string) axion $X$; it has to be rational in order to 
be able to $SL(2,Z)$ transform to a frame where $X$ 
vanishes\footnote{Similarly one needs a rational axion to be able to
$SL(2,Z)$ rotate into a perturbative open string regime.}. 
This would imply that when $X$ is irrational one is forced to 
present the D3-brane theory in terms of an $SL(2,Z)$ doublet of field 
strengths. This is just an observation, it is not clear to us at this 
time whether this has any profound meaning.

There are many extensions of our work that one can think of. 
First of all, the precise decoupling limit describing the 
non-commutative $(p,q)$ open strings in this context still has to
be worked out. Doing this is expected to connect to
(some of) the results reported in \cite{Gran-Nielsen}, either by
using a `flat space' closed string moduli approach or by making use
of holographic methods (using the dual supergravity background in order
to find a decoupling limit) \cite{Berman-HNC}. A very concrete goal
would for example be to obtain the $SL(2,R)$ invariant tension formula 
for the decoupled non-commutative open $(p,q)$ strings on the D3-brane. 
Another interesting project would be to see whether we can generalize
our discussion of $SL(2,R)$ covariant moduli of open strings on the 
D3-brane to the worldvolume of $(m,n)$ five-branes  
in Type IIB theory \cite{Gran-Nielsen}. This will inevitably introduce
a `little (closed) strings' sector into the theory and it would be 
interesting to see how this would affect our results. At first sight it 
seems like a good idea to take an OM-theory or M5-brane perspective 
again, because little strings can be naturally understood as open membranes
wrapped around a compact direction transverse to the 
M5-brane (this idea was first exploited in \cite{Larsson-Sundell}). 

From another point of view it would be very nice if one could provide 
a more direct understanding of the open membrane metric. So instead
of defining this object indirectly via string theory, we would prefer 
to understand the open membrane metric directly from an M-theory perspective. 
Quantizing the open membrane presumably has all the usual problems, 
so that does not seem to help us. A more fruitful, less ambitious, 
point of view seems to be the conformal relation of the open membrane 
metric to the M5-brane Boillat metric \cite{Gibbons-West}. This 
relation in some sense leaves us with the `smaller' problem of 
`explaining' the (rather complicated) conformal factor of the open membrane 
metric (\ref{OMexpr}) as compared to the M5-brane Boillat metric (where
the M5-brane Boillat metric can be understood as the metric 
naturally preferred by the M5-brane low energy effective equations 
of motion and providing the propagation cone for all perturbative degrees 
of freedom). This remains an important problem for the future.
 
\vspace{0.5cm}
{\bf \noindent Note added:} 
\vspace{0.5cm}

During the completion of this paper the preprint 
\cite{Berman-AdSOM} appeared, discussing an interesting $AdS_3$ 
self-dual string phase of OM-theory. In this preprint a conformal factor of 
the open membrane metric was presented that differed from ours. More recently 
this issue was resolved when the authors of \cite{Berman-AdSOM} corrected 
their conformal factor, obtained using a different method, which now
agrees with our result.

\acknowledgments
It is a pleasure to thank Jianxin Lu, Eric Bergshoeff, David Berman and 
especially Per Sundell for interesting discussions. The Groningen 
Institute for Theoretical Physics is acknowledged for their hospitality 
during which part of this work was done.

\bibliography{ometric}

\end{document}